\documentclass[final]{cvpr}
\usepackage[T1]{fontenc}
\usepackage{amsmath, amssymb}
\usepackage{newtxtext, newtxmath}
\usepackage{graphicx}
\usepackage[english]{babel}
\usepackage{epsfig}
\usepackage{makecell}
\usepackage{subfig}
\usepackage{fancyhdr}
\usepackage{booktabs}
\usepackage{adjustbox}
\usepackage[inline]{enumitem}
\usepackage{tabularx}
\usepackage{xcolor}
\usepackage{listings}
\usepackage{microtype}
\usepackage{etoolbox}
\usepackage{comment}
\usepackage{setspace}
\usepackage{etoolbox}
\usepackage{expl3}
\usepackage{bm}
\usepackage{xparse}

\makeatletter
\@namedef{ver@everyshi.sty}{}
\makeatother
\usepackage[skins]{tcolorbox}
\usepackage{tikz}

\usepackage{caption}
\usepackage{flushend}
\usepackage{multirow}

\usepackage[frozencache,cachedir={minted}]{minted}

\usepackage[inkscape=false]{svg}

\svgpath{{fig/}}

\usepackage[pagebackref=true,breaklinks=true,colorlinks,bookmarks=false]{hyperref}

\tcbuselibrary{listings, minted}
\usetikzlibrary{positioning, fit, calc, decorations.pathreplacing}
\pgfdeclarelayer{bg}
\pgfsetlayers{bg,main}

\let\vrb\lstinline

\makeatletter
\ExplSyntaxOn

\newcommand{\TikzAnchorLength}[5]{
	\pgfpointdiff{\pgfpointanchor{#1}{#2}}%
	             {\pgfpointanchor{#3}{#4}}%
	\expandafter\edef\csname #5\endcsname{\fp_eval:n {sqrt(\pgf@x*\pgf@x + \pgf@y*\pgf@y)}}
}

\box_new:N \l_sh_box
\dim_new:N \l_sh_dim
\dim_new:N \l_sd_dim

\newcommand{\bybox}[1]{
	\hbox_set:Nn \l_sh_box {\ttfamily M}
	\dim_set:Nn \l_sh_dim {\box_ht:N\l_sh_box}
	\adjustbox{fbox=0.2pt~0.4pt~0.5pt,raise=0.5pt}{\rule[0pt]{0pt}{1.2\l_sh_dim}#1}
}

\fp_new:N \l_my_fs_fp
\fp_new:N \l_my_bls_fp

\newcommand{\relsize}[1]{
    \fp_set:Nn \l_my_fs_fp {\f@size}
    \fp_set:Nn \l_my_bls_fp {\baselineskip}
    \exp_args:Nxx \fontsize{
        \fp_eval:n {(#1) * \l_my_fs_fp}
    }{
        \fp_eval:n {(#1) * \l_my_bls_fp}
    }
    \selectfont
}

\ExplSyntaxOff
\makeatother

\newcommand{\rev}[1]{#1}

\newcommand{\tablesize}{
    \fontsize{7}{7}\selectfont
}

\makeatletter
\g@addto@macro{\UrlBreaks}{\UrlOrds}
\makeatother

\setlist{left=0.1em,itemsep=0.0mm, parsep=1mm}

\def\cvprPaperID{35} 

\def\confYear{CVPR 2021}
\pretocmd{\paragraph}{\vspace*{-4.5mm}\relax}{}{
    \GenericError{unable to patch paragraph}{}{}{}
}

\makeatletter
\patchcmd{\@maketitle}{\@author}{\vspace*{-0.5mm}\@author}{}{\GenericError{failed to patch}{}{}{}}
\patchcmd{\@maketitle}{\vspace*{12pt}}{\vspace*{6pt}}{}{\GenericError{failed to patch}{}{}{}}
\patchcmd{\@maketitle}{\vspace*{24pt}}{\vspace*{18pt}}{}{\GenericError{failed to patch}{}{}{}}
\patchcmd{\@maketitle}{\vskip .5em}{}{}{\GenericError{failed to patch}{}{}{}}
\patchcmd{\cvprsection}{{10pt plus 2pt minus 2pt}{7pt}}{{5pt plus 1pt minus 1pt}{5pt}}{}{\GenericError{failed to patch}{}{}{}}
\patchcmd{\cvprsubsection}{{8pt plus 2pt minus 2pt}{6pt}}{{3pt plus 1pt minus 1pt}{3pt}}{}{\GenericError{failed to patch}{}{}{}}
\patchcmd{\abstract}{12pt}{8pt}{}{\GenericError{failed to patch}{}{}{}}
\patchcmd{\endabstract}{12pt}{7pt}{}{\GenericError{failed to patch}{}{}{}}
\makeatother

\begin{document}

\setlength{\belowdisplayskip}{1mm} \setlength{\belowdisplayshortskip}{0mm}
\setlength{\abovedisplayskip}{0mm} \setlength{\abovedisplayshortskip}{0mm}

\title{Forensic Analysis of Video Files Using Metadata}

\newcommand{\AuthorMarkA}{\textsuperscript{1}}
\newcommand{\AuthorMarkB}{\textsuperscript{2}}
\author{

{Ziyue Xiang$^{\star}$ \quad J\'{a}nos Horv\'{a}th$^{\star}$ \quad Sriram Baireddy$^{\star}$ }\\ {Paolo Bestagini$^{\dagger}$ \quad Stefano Tubaro$^{\dagger}$ \quad Edward J. Delp$^{\star}$}\\
    {\normalsize
    $^{\star}$ Video and Image Processing Lab (VIPER), School of Electrical and Computer Engineering,}\\
    {\normalsize Purdue University, West Lafayette, Indiana, USA}\\
    {\normalsize $^{\dagger}$ Dipartimento di Elettronica, Informazione e Bioingegneria, Politecnico di Milano, Milano, Italy
    }
}
\maketitle
{

\adjustbox{raise=7cm, set height=0mm, margin=3mm 0mm 0mm 0mm}{
  \parbox{1.8\linewidth}{
    \scriptsize\color{gray}
    \textcopyright 2021 IEEE.  Personal use of this material is permitted.  Permission from IEEE must be obtained for all other uses, in any current or future media, including reprinting/republishing this material for advertising or promotional purposes, creating new collective works, for resale or redistribution to servers or lists, or reuse of any copyrighted component of this work in other works.
  }
}

%
\begin{abstract}
The unprecedented ease and ability to manipulate video content has led to a rapid spread of manipulated media. 
The availability of video editing tools greatly increased in recent years, allowing one to easily generate photo-realistic alterations.
Such manipulations can leave traces in the metadata embedded in video files.
This metadata information can be used to determine video manipulations, brand of video recording device, the type of video editing tool, and other important evidence.
In this paper, we focus on the metadata contained in the popular MP4 video wrapper/container. 
We describe our method for metadata extractor that uses the MP4's tree structure.
Our approach for analyzing the video metadata produces a more compact representation. 
We will describe how we construct features from the metadata and then use
dimensionality reduction and nearest neighbor classification for forensic analysis of a video file. 
Our approach allows one to visually inspect the distribution of metadata features and make decisions.
The experimental results confirm that the performance of our approach surpasses other methods.
\end{abstract}

\section{Introduction}

The proliferation of easy-to-use video manipulation tools has placed unprecedented power in the hands of individuals.
Recently, an Indian politician used deepfake technology to rally more voters~\cite{charlotte_2020}.
In the original video the politician delivered his message in English; it was convincingly altered to show him speaking in local dialect.
Media manipulation methods are also used as tools of criticism and to undermine the reputation of politicians~\cite{raphael_2019}.
Such manipulated videos can now be easily generated to bolster disinformation campaigns and sway the public opinion on critical issues.

A wide variety of tools for video forensic analysis have been developed~\cite{milani_2012}. 
These tools can be used to attribute a video to the originating device, to reconstruct the past video compression history, and even to detect video manipulations. 
The most popular video manipulation detection techniques focus on inconsistencies and artifacts in the pixel domain~\cite{bayram_2008,koopman_2018,huy_2019,sabir_2019}.
As video manipulation detection methods become more sophisticated, video editing techniques continue to improve, leading to a situation where manipulated videos are becoming practically indistinguishable from real videos~\cite{brundage2018,ethics_2020,grobman_2019,ruiz_2020,umawing_2020,webb_2020}.
For this reason, detection techniques exploiting pixel-level analysis may fail, while methods that do not use pixel data will increasingly gain importance.

Video forensic techniques not exploiting pixel analysis typically work due to the presence of ``metadata''~\cite{Guera2019_ICMLW,iuliani_2019}.
This is additional embedded information that every video file contains.
The metadata are used for video decoding~\cite{keith_2007} and indicating other information such as the date, time, and location of the video when created. 
Because video editing tools tend to cause large structural changes in metadata, it is difficult for one to alter a video file without leaving any metadata traces. 
Therefore, metadata can serve as strong evidence in video forensics tasks.



In this paper, we leverage the seminal work presented in~\cite{iuliani_2019, yang_2020} to investigate the use of metadata for video forensic analysis of the MP4 and MOV video formats, which are among the most popular video wrappers/containers.
The MP4 format is used by numerous Android devices, social networks, and digital video cameras~\rev{\cite{android_formats, sony_camera_spec,social_network_video_spec}}. 
MOV format is mostly used by Apple devices and is \rev{derived from the same ISO standard as MP4~\cite{iso_media_part12}}. 
The design of the MP4 format is based \rev{on} MOV~\cite{quicktime_format_old}. 
The two formats can be parsed in a similar manner, thus we will refer to MP4 containers hereinafter even if MOV containers are considered.
As a result, our approach can analyze a large number of videos in the real world.

In our work, we examine the results of using the metadata in MP4 files for video forensic scenarios, extending the work presented in~\cite{yang_2020}. 
More specifically, we describe a metadata extraction method and improve the feature representation format to make metadata-based forensic feature vectors more compact. 
We employed feature selection techniques to boost the quality of the feature vectors. 
Finally, we reduced the dimensionality of the feature vectors to two, which allows visualization and classification in 2D space. 
We show that these feature vectors can be used for a wide variety of video forensic tasks, from source attribution to tampering detection.
Compared to other work, our proposed approach can generate 2D feature scatter plots and decision boundary graphs for many video forensics tasks. 
This feature enables us to gain insights into the distribution of MP4 metadata and make interpretable decisions. 

Our experimental results show that many video forensics problems on standard datasets can be solved reliably by looking only at metadata. 
We also discovered that videos uploaded to specific social networks (e.g., TikTok, WeiBo) present altered metadata, which invalidates metadata-based forensics methods. 
This is one of the limitations of our techniques and will be addressed in future research.

\section{Related Work}
Many techniques have been described to determine whether some part of a video has been tampered or not~\cite{milani_2012}.
Most of these methods were developed to detect manipulations in the pixel domain and do not use metadata information.
Compared to pixel-level analysis, metadata-based methods possess unique advantages. 
The size of metadata is significantly smaller than pixel data, which enables the analysis of large datasets in short amounts of time.
Most video manipulation software do not allow users to alter metadata directly~\cite{Guera2019_ICMLW, iuliani_2019}.
Consequently, metadata has a higher degree of opacity than pixel data, which makes metadata-based media forensics more reliable and its corresponding attacks more challenging.

Most existing work focuses on the metadata in MP4-like video containers, which maintain data in a tree structure. 
In \cite{iuliani_2019} and \cite{yang_2020}, the authors design features based on symbolic representation of MP4's tree structure, which are processed by a statistical decision framework and decision trees, respectively. 
The authors report high performance for both video authentication and attribution tasks. 
G\"{u}era \etal~\cite{Guera2019_ICMLW} extract video metadata with the \lstinline|ffprobe| utility and then do video authentication with an ensemble classifier. 

More low-level metadata-related information can be found by looking into video compression traces. 
Video compression  methods typically leave series of traces related to the way the video frames are compressed. 
This information is not easy to modify, thus acting as a metadata-like feature.
As an example, most contemporary video encoders compress frame sequences in a structures known as a Group of Pictures (GOP), where one frame can be defined using contents of other frames in order to save space~\rev{\cite{schafer_1995}}.
The dependency between frames within or across different GOPs may provide evidence for video manipulation.
Due to the complexity of video codecs, a number of techniques have been proposed for various settings of a codec where specific video encoding features are turned on or off. 
Vázquez-Padín \etal~\cite{padin_2020} provide a detailed explanation of the video encoding process and GOP structure.
They propose a video authentication method that generalizes across multiple video encoding settings.
Yao \etal~\cite{yao2020double} discuss the detection of double compression when an advanced video encoding feature called adaptive GOP is enabled.

\section{Proposed Approach}

\subsection{An Overview of Our Approach}
Video metadata captures multiple aspects of the history of a video file. 
In this paper we propose a framework that exploits an MP4 video file's metadata to solve multiple video forensics problems, including brand attribution, video editing tool identification, social network attribution, and manipulation detection. 
Our method can also be easily adapted to other forensics applications.

The structure of our proposed framework is illustrated in Figure~\ref{fig:framework-structure}. \rev{As will be discussed in Section \ref{subsec:mp4-metadata}, the MP4 format manages data using a tree structure.} First we extract the metadata from MP4 files while preserving their relationships in \rev{the} tree structure. The MP4 standard is around twenty years old, it contains numerous vendor-specific nuances that require separate parsing strategies. The metadata tree needs to go through several refining stages, which increase the granularity of the extracted information. In the next step, the tree representation of metadata is converted into a numeric \rev{feature} vector, which can be easily processed by machine learning methods. Our feature representation scheme is based upon \cite{yang_2020}. 
\rev{We improve the handling of media tracks and metadata fields that take on continuous values inside the tree.}
The resulting feature vectors preserve more characteristics of the videos, yet they tend to also be more compact. 
In the last stage, we use these features with a classifier based on the selected forensic application.
In the following we provide additional details about each step of our proposed framework.




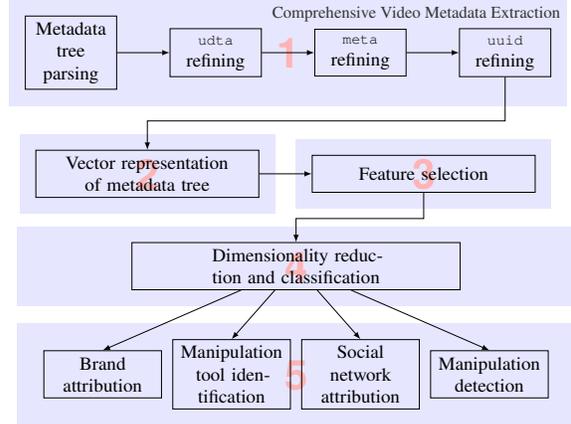
\begin{figure}[tb]
\centering
\tikzset{
	stepnd/.style={
		rectangle,
		draw,
		text width=1.5cm,
		text centered
	}
}

\ExplSyntaxOn
\newcommand{\QuickConsecutiveArrow}[3]{
	\int_step_inline:nnn {#2} {#3} {
		\int_compare:nNnT {##1} < {#3} {
			\draw[-latex] (#1##1) -- (#1\int_eval:n {##1 + 1});
		}
	}
}

\newcommand{\QuickSplitArrow}[3]{
	\clist_set:Nn \l_tmpa_clist {#3}
	\clist_map_inline:Nn \l_tmpa_clist {
		\draw[-latex] (#1#2)--(#1##1.north);
	}
}
\ExplSyntaxOff

\scalebox{0.7}{
\begin{tikzpicture}[node distance=1cm and 1cm]

\node[stepnd] (s1) {Metadata tree parsing};
\node[stepnd, right=of s1] (s2) {\vrb|udta|\\ refining};
\node[stepnd, right=of s2] (s3) {\vrb|meta|\\ refining};
\node[stepnd, right=of s3] (s4) {\vrb|uuid|\\ refining};
\begin{pgfonlayer}{bg}
\node[fit={(s1)(s2)(s3)(s4)}, fill=blue!10, inner sep=3mm, outer sep=0mm] (b1) {};
\node[anchor=north east] at (b1.north east) {\color{black!80}\small Comprehensive Video Metadata Extraction};
\node at (b1.center) {\Huge\bfseries\sffamily\color{red!30} 1};
\end{pgfonlayer}
\QuickConsecutiveArrow{s}{1}{4}

\node[stepnd, below left=of b1, text width=4cm, anchor=west, xshift=15mm, yshift=-3mm] (s5) {Vector representation of metadata tree};
\node[stepnd, right=of s5, text width=4cm, minimum height=2em] (s6) {Feature selection};

\coordinate (s5top) at ($(s5.north)+(0mm,5.5mm)$);
\draw[-latex] (s4) |- (s5top) -- (s5.north);
\QuickConsecutiveArrow{s}{5}{6}

\TikzAnchorLength{b1}{east}{b1}{west}{boxminwidth}

\begin{pgfonlayer}{bg}
\node[fit={(s5)}, fill=blue!10, inner sep=3mm, outer sep=0mm] (b2) {};
\node at (b2.center) {\Huge\bfseries\sffamily\color{red!30} 2};
\end{pgfonlayer}

\begin{pgfonlayer}{bg}
\node[fit={(s6)}, fill=blue!10, inner sep=3mm, outer sep=0mm] (b3) {};
\node at (b3.center) {\Huge\bfseries\sffamily\color{red!30} 3};
\end{pgfonlayer}

\node[stepnd, below=of s5, anchor=west, text width=6cm, yshift=-3mm, xshift=-3mm] (s7) {Dimensionality reduction and classification};
\coordinate (s7top) at ($(s7.north)+(0mm,4.5mm)$);
\draw[-latex] (s6) |- (s7top) -- (s7.north);

\begin{pgfonlayer}{bg}
\node[fit={(s7)}, fill=blue!10, inner sep=3mm, outer sep=0mm, minimum width=\boxminwidth] (b4) {};
\node at (b4.center) {\Huge\bfseries\sffamily\color{red!30} 4};
\end{pgfonlayer}

\begin{scope}[node distance=1cm and 2mm]
\node[stepnd, below left=of b4, yshift=-3mm, xshift=7mm, text width=2cm, anchor=west] (s8) {Brand attribution};
\node[stepnd, right=of s8, text width=2cm] (s9) {Manipulation tool identification};
\node[stepnd, right=of s9, text width=2cm] (s10) {Social network attribution};
\node[stepnd, right=of s10, text width=2cm] (s11) {Manipulation detection};
\end{scope}

\QuickSplitArrow{s}{7}{8,9,10,11}

\begin{pgfonlayer}{bg}
\node[fit={(s8)(s9)(s10)(s11)}, fill=blue!10, inner sep=3mm, outer sep=0mm, minimum width=\boxminwidth] (b4) {};
\node at (b4.center) {\Huge\bfseries\sffamily\color{red!30} 5};
\end{pgfonlayer}

\end{tikzpicture}
}
\caption{The structure of our proposed metadata forensic analysis technique. 
}
\label{fig:framework-structure}
\end{figure}

\subsection{Video Metadata}\label{subsec:mp4-metadata}
The first step in our approach consists of parsing metadata from video files.
Digital video files are designed to handle multimodal data such as video, audio, and subtitles.
The standards of these data modalities also shifts as technology advances. 
For example, since MP4's introduction in 2001, the mainstream video codec has changed from MPEG-2 to H.264, and may change to H.265 in the near future \rev{\cite{iso_generic_media,sullivan_2012,sullivan_2005}}. 
The metadata surrounding these various data modalities and standards are inserted into a video file in distinct ways. In this paper, we use the word \emph{comprehensive} to describe a video metadata extraction scheme that is capable of parsing metadata across most data modalities and encoding specifications.

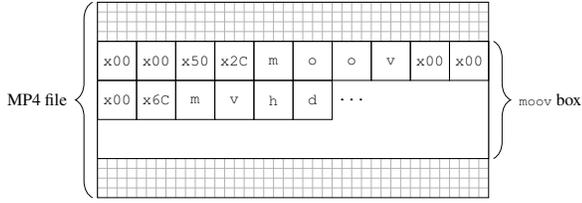
\begin{figure}[tb]
\centering
\ExplSyntaxOn

\int_new:N \l_num_cols_int
\int_new:N \l_num_rows_int
\tl_new:N \l_index_tl
\tl_new:N \l_cell_text_tl
\int_set:Nn \l_num_cols_int {10}
\int_set:Nn \l_num_rows_int {5}
\dim_new:N \l_cell_w_dim
\dim_new:N \l_cell_h_dim
\dim_set:Nn \l_cell_w_dim {8mm}
\dim_set:Nn \l_cell_h_dim {8mm}
\fp_new:N \l_grid_factor_fp
\int_new:N \l_grid_factor_int
\int_set:Nn \l_grid_factor_int {4}

\tikzset{
	cellnd/.style={
		minimum~height=\l_cell_h_dim,
		minimum~width=\l_cell_w_dim,
		inner~sep=0pt,
		outer~sep=0pt,
		draw=black,
		text~centered,
		execute~at~begin~node=\ttfamily\small
	}
}

\prop_new:N \l_cell_text_prop
\prop_set_from_keyval:Nn \l_cell_text_prop {
	1=x00,
	2=x00,
	3=x50,
	4=x2C,
	5=m,
	6=o,
	7=o,
	8=v,
	9=x00,
	10=x00,
	11=x00,
	12=x6C,
	13=m,
	14=v,
	15=h,
	16=d
}

\cs_set:Npn \cell_loc:nn #1#2 {
	\fp_eval:n {(#2 - 1)*\l_cell_h_dim} pt, \fp_eval:n {-(#1 - 1)*\l_cell_w_dim} pt
}

\cs_set:Npn \cell_loc:nnnn #1#2#3#4 {
	\fp_eval:n {(#2 - 1)*\l_cell_h_dim/(#4)} pt, \fp_eval:n {-(#1 - 1)*\l_cell_w_dim/(#3)} pt
}

\cs_set:Npn \draw_box:nnn #1#2#3 {
	\node[cellnd, 
	minimum~width=\fp_eval:n {\l_num_cols_int * \l_cell_w_dim}pt,
	minimum~height=\fp_eval:n {(#2-(#1)+1) * \l_cell_h_dim},
	anchor=north~west, #3
	]
	at 
	(\cell_loc:nn {#1}{1}) {};
}

\tl_new:N \l_tmpc_tl

\cs_set:Npn \draw_background:nn #1#2 {
	\edef\l_tmpa_tl{\int_eval:n {(#2-(#1) + 1) * \l_grid_factor_int}}
	\edef\l_tmpb_tl{\int_eval:n {\l_num_cols_int * \l_grid_factor_int}}
	
	\int_step_inline:nnn {(#1 - 1) * \l_grid_factor_int + 1} {(#1 - 1) * \l_grid_factor_int + \l_tmpa_tl} {
		\draw[color=black!30] (\cell_loc:nnnn{##1}{1}{\l_grid_factor_int}{1})
		--
		(\cell_loc:nnnn{##1}{\l_num_cols_int + 1}{\l_grid_factor_int}{1});
	}
	
	\int_step_inline:nnn {1} {\l_tmpb_tl} {
		\draw[color=black!30] (\cell_loc:nnnn{#1}{##1}{1}{\l_grid_factor_int})
		--
		(\cell_loc:nnnn{#2 + 1}{##1}{1}{\l_grid_factor_int});
	}
}

\scalebox{0.65}{
\begin{tikzpicture}

\draw_background:nn{0}{4}
\draw_box:nnn {0}{4}{}
\draw_box:nnn {1}{3}{fill=white}

\int_step_variable:nNn {\l_num_rows_int} \l_tmpa_tl {
	\int_step_variable:nNn {\l_num_cols_int} \l_tmpb_tl {
		\edef\l_index_tl{\int_eval:n {(\l_num_cols_int) * (\l_tmpa_tl - 1) + \l_tmpb_tl}}
		\prop_get:NVNT \l_cell_text_prop \l_index_tl \l_cell_text_tl {
			\node[cellnd,anchor=north~west]
			at (\cell_loc:nn {\l_tmpa_tl}{\l_tmpb_tl}) 
			{\l_cell_text_tl};
		}
	}
}

\node[cellnd, draw=none, anchor=north~west] at (\cell_loc:nn {2}{7}) {\ldots};
\coordinate (mp4t) at (\cell_loc:nn {0}{1});
\coordinate (mp4b) at (\cell_loc:nn {5}{1});

\coordinate (moovt) at (\cell_loc:nn {1}{11});
\coordinate (moovb) at (\cell_loc:nn {4}{11});

\draw [decorate,decoration={brace,amplitude=10pt,mirror}, transform~canvas={xshift=-1mm}]
(mp4t) -- (mp4b) node [black,midway,anchor=east,xshift=-4mm] {MP4~file};
\draw [decorate,decoration={brace,amplitude=10pt}, transform~canvas={xshift=1mm}]
(moovt) -- (moovb) node [black,midway,anchor=west,xshift=4mm] {\vrb|moov|~box};

\end{tikzpicture}
}
\ExplSyntaxOff

\caption{Illustration of the MP4 file format, where each cell represents one byte. An MP4 file is made up of a series of \emph{boxes}. Every box has an 8-byte header, where the first 4 bytes store the size of the box in bytes (including the header) as a 32-bit big-endian integer, and the last 4 bytes store the name of the box. The content of a box is either child boxes or binary data. Binary data from different boxes may require distinct decoding methods.}
\label{fig:mp4-structure}
\end{figure}

Figure~\ref{fig:mp4-structure} shows the basic structure of an MP4 file \rev{\cite{iso_mp4}}. 
An MP4 file is composed by a number of \emph{boxes}. Each box starts with a header that specifies the box size (in byte) and name. 
The size helps the user to locate box boundaries, and the name defines the purpose of the box. The content of a box can either be a child box or some encoded binary data. 
Thus, an MP4 file is effectively a tree structure, where information is stored on leaf nodes. 
Given MP4's tree structure, we can capture metadata information at two levels:
\begin{enumerate*}[label=(\arabic*)]
\item the structure of the tree; and
\item the interpreted values of binary data contained in leaf nodes
\end{enumerate*}. 
Therefore, the job of a metadata extractor is to traverse the MP4 tree, attain the tree's structure, filter non-metadata leaf nodes (e.g., nodes that contain video compressed pixel information) and interpret values on relevant leaves. 
As shown in Figure~\ref{fig:tree-string}, the output of a metadata extractor can be represented without any loss by a collection of human-readable strings.

\begin{figure}[tb]
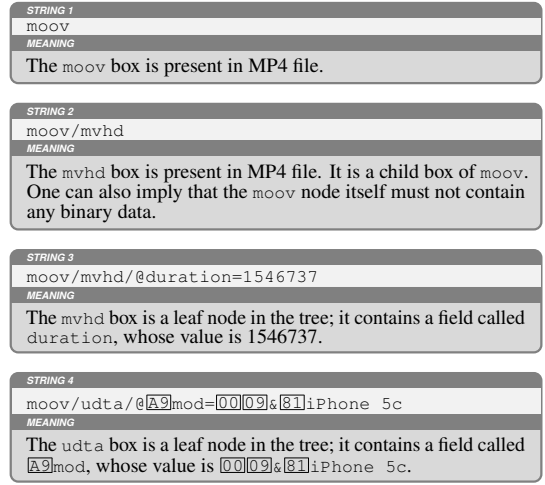

\centering

\newcommand\stringtitlestyle{\tiny\sffamily\bfseries\itshape\color{white}}
\newcommand{\customboxlower}{
\tcblower
\vspace*{3mm}
}

\ExplSyntaxOn

\int_new:N \g_string_box_int
\int_gset:Nn \g_string_box_int {0}

\tcbset{
	stringbox/.style={
		bicolor,
		left=2mm,
		right=2mm,
		top=0.5mm,
		middle=0.25mm,
		bottom=0.5mm,
		boxrule=1pt,
		scale=0.85,
		title={\stringtitlestyle STRING~\int_use:N \g_string_box_int},
		colframe=black!50,
		overlay={
			\TikzAnchorLength{segmentation}{west}{segmentation}{east}{boxsegwidth}
			\edef \l_tmpa_tl {\fp_eval:n {\boxsegwidth + 0.5}}
			\coordinate (segw) at ($(segmentation.west)+(-0.5pt,0.5pt)$);
			\node[fill=black!50, 
			anchor=north~west, 
			minimum~width=\l_tmpa_tl,
			inner~sep=1.8pt,
			align=left
			] 
			at (segw) {\phantom{\stringtitlestyle M}};
			\node[anchor=north~west, 
			outer~sep=0pt, 
			inner~sep=0pt,
			xshift=2.5mm,
			yshift=-0.75mm
			] at (segw) {\stringtitlestyle MEANING};
		},
		boxsep=1pt,
		bottomtitle=0pt,
		code={\int_gincr:N\g_string_box_int\setstretch{0.3}\small}
	}
}
\ExplSyntaxOff

\begin{tcolorbox}[stringbox]
\vrb|moov|
\customboxlower

The \vrb|moov| box is present in MP4 file.
\end{tcolorbox}

\begin{tcolorbox}[stringbox]
\vrb|moov/mvhd|
\customboxlower
The \vrb|mvhd| box is present in MP4 file. It is a child box of \vrb|moov|. One can also imply that the \vrb|moov| node itself must not contain any binary data.
\end{tcolorbox}

\begin{tcolorbox}[stringbox]
\vrb|moov/mvhd/@duration=1546737|
\customboxlower
The \vrb|mvhd| box is a leaf node in the tree; it contains a field called \vrb|duration|, whose value is 1546737. 
\end{tcolorbox}

\begin{tcolorbox}[stringbox]
\vrb|moov/udta/@$\bybox{A9}$mod=$\bybox{00}\bybox{09}$&$\bybox{81}$iPhone 5c|
\customboxlower
The \vrb|udta| box is a leaf node in the tree; it contains a field called \vrb|$\bybox{A9}$mod|, whose value is \vrb|$\bybox{00}\bybox{09}$&$\bybox{81}$iPhone 5c|. 
\end{tcolorbox}

\caption{Examples of representing MP4 metadata tree with strings. Node paths are separated with `\protect\vrb|/|', the values of leaf nodes are prefixed with `\protect\vrb|@|', non-ASCII and unprintable characters are shown as hexadecimal codes surrounded by black frames. The metadata tree of any MP4 file can be portrayed by a collection of such strings.}
\label{fig:tree-string}
\end{figure}


Our metadata extractor focuses on vendor-specific non-standard MP4 metadata boxes that are significant for forensic purposes.
More precisely, we determine that \vrb|udta|, \vrb|uuid|, \vrb|meta|, and \vrb|ilst| boxes are likely to carry vital information for video forensics. 
We next discuss our strategies to refine the parsing process. 

\subsubsection{Parsing \protect\vrb|ilst| Data}

\vrb|ilst| (``metadata \underline{i}tem \underline{l}i\underline{st}'') boxes in MP4 files are used to store vendor-specific metadata \cite{qt_format_spec}. 
Generally speaking, \vrb|ilst| boxes are container boxes whose child boxes carry metadata items as key-value pairs. 
The names of \vrb|ilst|'s children (i.e., the \emph{keys}) would start with \vrb|$\bybox{A9}$| (equivalent to character `\textcopyright'). 
A list of frequently used \vrb|ilst| keys is shown in Table \ref{table:ilst-keys}. 
One can see that the content of the \vrb|ilst| box is particularly important for forensic analysis, for it often contains information about the manufacturer of the video capturing device, the encoder version, and the location and time of the capture. 

\begin{table}[b]
\tablesize
\centering
\caption{A list of common keys in \protect\vrb|ilst| boxes \cite{qt_format_spec_moov}.}
\begin{tabular}{c>{\centering\arraybackslash}p{0.6\linewidth}}
\toprule
Key & Description\\ \midrule
\vrb|$\bybox{A9}$mod| & camera model\\
\vrb|$\bybox{A9}$too| & name and version of encoding tool\\
\vrb|$\bybox{A9}$nam| & title of the content\\
\vrb|$\bybox{A9}$swr| & name and version number of creation software\\
\vrb|$\bybox{A9}$swf| & name of creator\\
\vrb|$\bybox{A9}$day| & timestamp of the video\\
\vrb|$\bybox{A9}$xyz| & geographic location of the video\\
\bottomrule
\end{tabular}
\label{table:ilst-keys}
\end{table}

It is difficult to parse the \vrb|ilst| box because various device manufacturers employ vastly different approaches when using it. 
Below, we report some interesting variants we found during our experiments:

\begin{itemize}
\item \vrb|ilst|'s child boxes directly placed in \vrb|moov/udta|

In some old Apple devices (e.g., iPhone 4, iPhone 5c, iPad 2), the child boxes of \vrb|ilst| are placed directly in \vrb|moov/udta| box. 

\item \vrb|ilst| boxes in \vrb|moov/meta|

As its name suggests, the \vrb|meta| box is used to store metadata. 
In this case, the \vrb|meta| box behaves similarly as other standard boxes, which means it can be parsed normally. 
As the MP4 parser traverses the box, it will eventually reach \vrb|ilst| and its children.

\item \vrb|ilst| boxes in \vrb|moov/udta/meta| and \vrb|moov/trak/meta|

When \vrb|meta| boxes appear in \vrb|udta| and \vrb|trak| boxes, they deviate from standard boxes. 
More specifically, 4 extra bytes are inserted right after the \vrb|meta| header to store information~\cite{mp4_meta_breakdown}. 
These types of \vrb|meta| boxes cannot be parsed by the MP4 parser, normally because the program will see these 4 bytes as the size of next box, which will lead to corrupted results. 

\end{itemize}

Our comprehensive metadata extractor is able to distinguish between these three scenarios and process MP4 video files correctly by fine-tuning the parsing of \vrb|udta| and \vrb|meta| boxes.

\subsubsection{Parsing XML Data}

We concluded that many video files contain XML data after inspecting numerous video files, especially those edited by ExifTool and Adobe Premiere Pro. 
These tools make use of the Extensible Metadata Platform (XMP) to enhance metadata management, which introduces a large amount of metadata inside an MP4 file's \vrb|uuid| and \vrb|udta| boxes in the form of XML. 
In Figure \ref{fig:xml-example}, we show two XML examples extracted from MP4 containers. It can be seen that these XML data can potentially contain a large amount of information, which includes type of manipulation, original value before manipulation, and even traces to locate the person that applied the manipulation. 
It is vital for our metadata extractor to have the ability to handle XML data inside MP4 files.

To avoid over-complicating the extracted metadata tree, we discard the tree structure of XML elements and flatten them into a collection of key-value pairs. 
If there is a collision between key names, the older value will be overwritten by the newer one, which indicates that only the last occurring value of each key is preserved. 
Despite the fact that some information is lost by doing so, the data we have extracted is likely to be more than enough for most automated video forensic applications.

\begin{figure*}[ht]

\tcbset{
	xmlisting/.style={
	listing only,
	listing engine=minted,
	minted language=HTML,
	minted options={
		breaklines,
		obeytabs,
		fontsize=\fontsize{5}{5}\selectfont,
		breakafter={abcdefghijiklmnopqrstuvwxyz1234567890}
	}
	},
	left=1mm,
	right=1mm,
	top=1mm,
	bottom=1mm,
	boxrule=0.4pt,
	width=0.45\textwidth,
	valign=top,
	colback=black!2
}

\centering
\subfloat[Excerpt of XML data from a video processed by Adobe Premiere Pro. It clearly contains multiple important timestamps, the software name and version, and even the path to the Premiere project.]
{\tcbinputlisting{xmlisting, listing file={listing/premiere.xml}}}
\qquad
\subfloat[XML data from a video modified by ExifTool. It can be seen that the version of ExifTool is 11.37; the presence of \protect\vrb|exif:DateTimeOriginal| implies the date of the video is modified.]
{\tcbinputlisting{xmlisting, listing file={listing/exiftool.xml}}}
\par

\caption{Examples of XML data in MP4 video containers. }
\label{fig:xml-example}
\end{figure*}


\subsection{Track and Type Aware Feature}
The second step of our approach consists of turning the parsed metadata into feature vectors.
{Most machine learning methods use vectors as the input. The string representation of metadata trees generated in the previous step needs to be transformed into feature vectors before being used by machine learning methods.}

\newcommand{\cate}[1]{(#1)}
\newcommand{\cateset}[1]{C_{\cate{#1}}}

\NewDocumentCommand{\catenum}{om}{
	\IfValueTF{#1}{
		\left\lvert\cateset{#2}^{#1}\right\rvert%
	}{
		\left\lvert\cateset{#2}\right\rvert%
	}
}

\NewDocumentCommand{\featind}{O{1}om}{%
	\IfValueTF{#2}{
		\adjustbox{raise=1pt,scale={1.0}{1.1}}{$\chi$}_{\cate{#1}}^{#2}(#3)%
	}{
		\adjustbox{raise=1pt,scale={1.0}{1.1}}{$\chi$}_{\cate{#1}}(#3)%
	}
}

\newcommand{\vecele}[2]{\bm{#1}_{\left[#2\right]}}

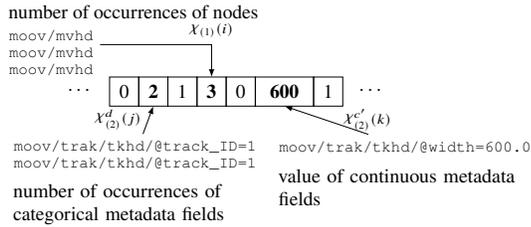
\begin{figure}[tb]
\centering
\tikzset{
	metand/.style={
		rectangle
	},
    vecnd/.style={
        rectangle,
        draw,
        inner sep=0pt,
        outer sep=0pt,
        minimum height=1.5em,
        text centered
    }
}

\ExplSyntaxOn

\dim_new:N \l_my_bw_dim
\dim_set:Nn \l_my_bw_dim {1.5em}

\dim_new:N \l_my_x_dim
\dim_gset:Nn \l_my_x_dim {0pt}

\int_new:N \l_my_vec_int
\int_gset:Nn \l_my_vec_int {1}

\cs_set:Nn \__my_node:nnnn {
    \node[#1] (#2) at (#3) {#4};    
}

\cs_generate_variant:Nn \__my_node:nnnn {nVnn}

\NewDocumentCommand{\vecnode}{O{1.0}O{0pt}mO{}}{
    \edef \l_tmpa_tl {vec\int_use:N \l_my_vec_int}
    \__my_node:nVnn {vecnd, minimum~width=\dim_eval:n {#1\l_my_bw_dim}, anchor=west, #4}
    \l_tmpa_tl {\dim_use:N \l_my_x_dim, #2} {#3}
    \dim_add:Nn \l_my_x_dim {#1\l_my_bw_dim}
    \int_gincr:N \l_my_vec_int
}

\ExplSyntaxOff
\resizebox{0.9\linewidth}{!}{
\begin{tikzpicture}[node distance=3mm]

\node[metand] (m1) at (0, 1) {\vrb|moov/mvhd|};
\node[metand, below=of m1.west, anchor=west] (m2) {\vrb|moov/mvhd|};
\node[metand, below=of m2.west, anchor=west] {\vrb|moov/mvhd|};

\node[fit={(m1)(m2)}, outer sep=0mm, inner sep=0mm] (m_12) {};
\node[above left=of m_12, anchor=west, xshift=2mm] {number of occurrences of nodes};

\vecnode[2]{$\cdots$}[draw=none]
\vecnode{0}
\vecnode{\textbf{2}}
\vecnode{1}
\vecnode{\textbf{3}}
\vecnode{0}
\vecnode[2]{\textbf{600}}
\vecnode{1}
\vecnode[2]{$\cdots$}[draw=none]

\draw[-latex] (m_12)-| (vec5) node [midway, above] {\scriptsize $\featind{i}$};

\node[metand, anchor=west, xshift=-8mm] (m3) at (0, -1) {\vrb|moov/trak/tkhd/@track_ID=1|};
\node[metand, below=of m3.west, anchor=west] (m4) {\vrb|moov/trak/tkhd/@track_ID=1|};
\node[fit={(m3)(m4)}, outer sep=0mm, inner sep=0mm] (m_34) {};
\node[below left=of m_34, anchor=west, text width=4cm, yshift=-3mm, xshift=2mm] {number of occurrences of categorical metadata fields};

\draw[-latex] (m_34)-- (vec3.south) node [midway, left] {\scriptsize $\featind[2][d]{j}$};

\node[metand, anchor=west] (m6) at (4, -1) {\vrb|moov/trak/tkhd/@width=600.0|};
\node[below left=of m6, anchor=west, text width=5cm, yshift=-3mm, xshift=2mm] {value of continuous metadata fields};
\draw[-latex] (m6)-- (vec7.south) node [midway, right, xshift=2mm] {\scriptsize $\featind[2][c^{\prime}]{k}$};

\end{tikzpicture}}

\caption{Illustration of the vector representation of MP4 metadata. The \protect\adjustbox{raise=1pt,scale={1.0}{1.1}}{$\chi$} functions help determine the corresponding element of a node or a metadata field in the feature vector.}
\label{fig:metadata-vec-repr}
\end{figure}

Our feature representation technique is shown in Figure \ref{fig:metadata-vec-repr}. For feature vectors to contain sufficient information of the MP4 tree, they need to include two levels of details: structure of the tree and value of metadata fields.
Metadata can be either categorical or continuous numerical fields. Considering categorical values, we assign each node and metadata key-value pair in the MP4 tree an element in the feature vector, which counts the number of occurrences of that node or pair. This strategy preserves information about the MP4 tree in the feature vector to a great extent. 
Considering numerical values, creating a new element for each of these key-value pairs will render the feature vectors large and redundant. We decide to insert the values into the feature vectors directly.

From Figure~\ref{fig:tree-string}, we know that string representation can be put into two categories:
\begin{enumerate*}[label=(\arabic*)]
\item strings that indicate the presence of a node, with node path separated by `\vrb|/|'; and
\item strings that show the key-value pair stored in a node, with node path separated by `\vrb|/|' and key-value separated by `\vrb|=|'.
\end{enumerate*}
Since an MP4 file can be seen as a collection of such strings, the feature transformation process can be viewed as a mapping from a given collection of strings $S$ to a vector $\bm{v}$. 
For the discussion below, we use $\vecele{v}{l}$ to denote the $l$-th element of $\bm{v}$.

In order to construct a mapping $S \rightarrow \bm{v}$, we need to consider the set of all possible strings $\Omega$. 
We denote the set of all category (1) strings and category (2) strings by $\cateset{1}$ and $\cateset{2}$, respectively. 
By definition, $\cateset{1}$ and $\cateset{2}$ form a partition of $\Omega$. 
We assume that both $\cateset{1}$ and $\cateset{2}$ are ordered so that we can query each element by its index. 
Let us denote the $l$-th elements of $\cateset{1}$ and $\cateset{2}$ by $\cateset{1}[l]$ and $\cateset{2}[l]$, respectively.

Each category (1) string corresponds to an element in $\bm{v}$. 
For the $i$-th category (1) string, we denote the index of the corresponding vector element in $\bm{v}$ by $\featind{i}$.
The value corresponding to this element is given by
\begin{multline}
\vecele{v}{\featind{i}} = \mbox{number of times $\cateset{1}[i]$ occurrs in $S$} , \\
\forall i \in \left\{1, 2, \ldots, \catenum{1}\right\}.
\end{multline}

We treat each media track (\vrb|trak|) segment in the metadata strings in a different way. In the MP4 file format, each \vrb|trak| node contains information about an individual track managed by the file container.
We observed that the number of tracks and content of the tracks remain consistent among devices of the same model. 
The structure of an MP4 file can be better preserved if we distinguish different tracks rather than merging them. 
This is achieved by assigning each track a track number in the metadata extraction phase. 
For example, the first track will be \vrb|moov/trak1|, and the second track will be \vrb|moov/trak2|. 
As a result, the child nodes and key-value pairs of each track will be separated, which effectively makes the feature vectors \emph{track-aware}.

For category (2) strings \rev{that represent key-value pairs stored in a node}, we applied a slightly different transformation strategy. 
We observed that some fields in MP4 files are essentially \emph{continuous} (e.g., \vrb|@avgBitrate|, \vrb|@width|). 
Despite the fact that most MP4 metadata fields are discrete, assigning each combination of these continuous key-value pairs a new element in $\bm{v}$ will still result in large and redundant feature vectors. 
We continue to subdivide $\cateset{2}$ based on the \emph{type} of each field, where the set of strings that have discrete fields is denoted by $\cateset{2}^d$, and the set of strings that have continuous fields is denoted by $\cateset{2}^c$. 
For strings that belong to $\cateset{2}^d$, the transformation scheme is similar to that of category (1) strings. 
Let the vector element index of the $j$-th string in $\cateset{2}^d$ be $\featind[2][d]{j}$, then the value corresponding to the element is given by
\begin{multline}
\vecele{v}{\featind[2][d]{j}} = \mbox{number of times $\cateset{2}^d[j]$ occurs in $S$} , \\
\forall j \in \left\{1, 2, \ldots, \catenum[d]{2}\right\}.
\end{multline}
As for strings that belong to $\cateset{2}^c$, we first discard the values in the strings to form a set of distinct keys $\cateset{2}^{c^{\prime}}$, and then put the values in $\bm{v}$ directly. 
For the $k$-th string in $\cateset{2}^{c^{\prime}}$, the index of the corresponding vector element in $\bm{v}$ is $\featind[2][c^{\prime}]{j}$, and the value of the element is
\begin{multline}
\vecele{v}{\featind[2][c^{\prime}]{k}} = 
\begin{cases}
\mbox{the value of key $\cateset{2}^{c^{\prime}}[k]$ in $S$} \\
0~\mbox{(if the key $\cateset{2}^{c^{\prime}}[k]$ is not in $S$)}
\end{cases}
 , \\
\forall j \in \left\{1, 2, \ldots, \catenum[c^{\prime}]{2}\right\}.
\end{multline}
It can be seen that the dimensionality of $\bm{v}$ is given by 
\begin{align}
\dim(\bm{v}) = \catenum{1}+\catenum[d]{2}+\catenum[c^{\prime}]{2}.
\end{align}
In general, the actual value of the index functions $\adjustbox{raise=1pt,scale={1.0}{1.1}}{$\chi$}$ can be arbitrary as long as all values of $\featind{i}$, $\featind[2][d]{j}$, and $\featind[2][c^{\prime}]{k}$ form a valid partition of integers in the range $\left[1, \dim(\bm{v})\right]$.

\rev{By using different representation strategies for discrete and continuous fields (i.e., being \emph{type-aware}), the resulting feature vectors are more compact and suited to machine learning techniques.}

\subsection{Feature Selection}
Our third step consists of reducing the set of selected features by discarding redundant features.
\rev{Based on the feature extraction scheme above, it can be observed that some elements in the feature vector are significantly correlated.}
For example, the presence of string \vrb|moov/mvhd/@duration=1546737| in a collection $S$ extracted from a valid MP4 file must lead to the presence of \vrb|moov/mvhd| in $S$. Therefore, feature selection can reduce redundancy within the feature vectors.


In the feature selection step, we reduce the redundancy among features that are strongly correlated. 
Since only a small proportion of elements in the feature vector $\bm{v}$ are inserted from continuous fields,
most elements in $\bm{v}$ correspond to the number of occurrences of a node or a field value. 
If two features in $\bm{v}$ are negatively correlated, then it often means the presence of an MP4 entity implies the absence of another MP4 entity. 
In forensic scenarios, presence is much stronger evidence than absence. 
Therefore, if we only focus on features that are positively correlated, then we can select features of higher forensic significance.

\rev{Given a set of feature vectors $\bm{v}_1, \bm{v}_2, \ldots, \bm{v}_N$, we can compute the corresponding correlation matrix $\bm{R}$, where $\bm{R}_{ij}$ is equal to the correlation coefficient between $i$-th and $j$-th feature in the set. Then, we set all negative elements in $\bm{R}$ to zero, which results in matrix $\bm{R}^+$.}
That is, negative correlation is ignored. 
\rev{Because all elements in $\bm{R}^+$ are within the range $[0,1]$, the matrix $\bm{R}^+$ can be seen as an affinity matrix between $\dim(\bm{v})$ vertices in a graph, where an affinity value of 0 indicates no connection and an affinity value of 1 indicates strongest connection.}
This allow us to use spectral clustering with $\alpha$ clusters on $\bm{R}^+$, which assigns multiple strongly correlated features into the same cluster \cite{shi2000normalized}. 
Then, we select clusters with more than $\beta$ features.
{For each selected cluster, we retain only one feature at random.}
Here, $\alpha > \beta > 0$ are hyperparameters. 
This feature selection step helps improve the quality of feature vectors. 

\subsection{Dimensionality Reduction and Classification}

In the last step, depending on the the video forensics problem, we use the feature vectors for classification in two ways.

\paragraph{Multi-class problems} When the classification problem is multi-class, we use linear discriminant analysis (LDA) \cite{friedman2001elements} to drastically reduce the dimensionality of feature vectors to 2 dimensions. LDA is a supervised dimensionality reduction technique.
For a classification problem of $K$ classes, LDA generates an optimal feature representation in a subspace that has dimensionality of at most $K-1$. 
The new features in the subspace are optimal in the sense that the variance between projected class centroids are maximized. 
For multi-class classification problems, we always reduce the dimensionality of the feature vector to 2.
After dimensionality reduction, we use a nearest neighbor classifier that uses the distance between the query sample and $\lambda$ nearest samples to make a decision, where $\lambda$ is a hyperparameter. 
Each nearest sample is weighted by their distance to the query sample. The use of the dimensionality reduction and nearest neighbor classifier lead to concise and straightforward decision rules in 2D space, which can be interpreted and analyzed by human experts easily.

\paragraph{Two-class problems} When the classification problem is two-class ($K=2$), LDA can only generate one-dimensional features. Our experiments have shown that 1D features are insufficient to represent the underlying complexity of video forensics problems. As a result, for binary classification problems, we use a decision tree classifier without dimensionality reduction.

\section{Experiments and Results}
In this section we report all the details of the experiments and comment on the results.


We study the effectiveness of our approach using the following datasets:

\begin{itemize}
    \item VISION \cite{shullani_2017}: the VISION dataset contains 629 pristine MP4/MOV videos captured by 35 different Apple, Android and Microsoft mobile devices.
    \item EVA-7K \cite{yang_2020}: the EVA-7K dataset contains approximately 7000 videos captured by various mobile devices, uploaded to distinct social networks, and edited by different video manipulation tools. The authors took a subset of videos from the VISION dataset and edited them with a number of video editing tools. Then, they uploaded both the original videos and edited videos to four social networks, namely YouTube, Facebook, TikTok and WeiBo. The videos were subsequently downloaded back. The EVA-7K dataset is made up of the pristine videos, edited videos, and downloaded videos.
\end{itemize}

The VISION dataset is used for device attribution experiments, while all other experiments are conducted on the larger EVA-7K dataset. 


We demonstrate the effectiveness of our approach in four video forensic scenarios. 
In all of the experiments below that use LDA and nearest neighbor classification, we choose $\alpha=300$, $\beta=4$, and $\lambda = 5$. 
For each of the scenarios below, unless indicated otherwise, the dataset is split into two halves with stratified sampling for training and validation, respectively. 
The metadata nodes and fields that are excluded during metadata extraction, as well as the list of continuous features in the vector representation step, are provided in supplementary materials. 
We mainly compare the performance of our method to \cite{yang_2020}, where the EVA-7K dataset is described. 
For brand attribution, because it is conducted on the VISION dataset, we select~\cite{iuliani_2019} as the baseline. 
The experiment results show that our approach achieves high performance evaluation metrics in all four scenarios. 

\subsection{Scenario 1: Brand Attribution}

Brand attribution consists of identifying the brand of the device used to capture the video file under analysis.
We examined brand attribution experiments in two scenarios.
In the first experiment, we assume the analyst has access to all possible devices at training time (i.e., close-set scenario).
In the second experiment, we assume that a specific device may be unknown at training time (i.e., blind scenario).

\paragraph{Close-set scenario}
In Table \ref{table:device-attribution-f1}, we show the F1-score comparison between our approach and previous work. 
Our method almost perfectly classifies the VISION dataset, with only one Apple device being misclassified as Samsung. 
{Because our framework is capable of extracting and analyzing more metadata, the performance of our method is better compared to the baseline, especially for brands like Huawei, LG, and Xiaomi.}
The 2D feature distribution and decision boundary for each label are shown in Figure \ref{fig:brand-attr-plot}, from which we can determine the metadata similarity between brands and the metadata ``variance'' for each brand. 
{Visualizations as shown in Figure 6 generated by our method can aid an analyst in making a more interpretable and reliable decision.
If a new video under examination is projected close to the LG region, one can assume it is unlikely for that video to come from a Samsung device.
}

\begin{figure}[htpb]
    \centering
    \includesvg[width=0.8\linewidth]{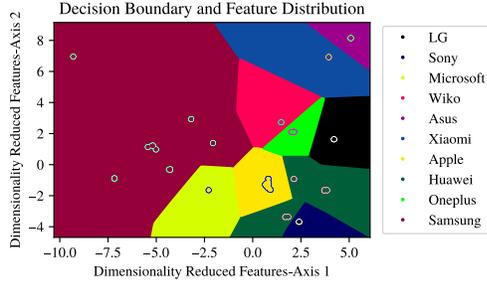}
    \caption{2D feature distribution and classification boundary for device attribution scenario.}
    \label{fig:brand-attr-plot}
\end{figure}

\begin{table}[htpb]
    \tablesize
    \centering
    \vspace*{2mm}
    \caption{F1-score comparison of device attribution scenario.}
    Device Attribution\\
    \begin{tabular}{rcc}
    \toprule
        \multicolumn{1}{c}{Brand} & Iuliani, \etal \cite{iuliani_2019} & Our Approach \\ \midrule
        Apple & 1.00  & 0.99 \\ 
        Asus & 1.00  & 1.00 \\ 
        Huawei &  0.87 & 1.00  \\ 
        LG & 0.94  & 1.00 \\ 
        Oneplus & 0.93  & 1.00 \\ 
        Samsung & 0.93  & 0.99 \\ 
        Wiko &  0.65 & 1.00 \\
        Xiaomi & 0.74 & 1.00 \\ \bottomrule
    \end{tabular}
    \label{table:device-attribution-f1}
\end{table}

\paragraph{Blind scenario}

We also examined device attribution with an unknown device using our approach. 
{Let us consider a specific example where device ID 20 (an Apple device) is the unknown device.}
In the training phase, we use the entire VISION dataset except for samples from device ID 20.
In order to classify this device that is unknown to the classifier, we project its features in 2D space and plot them in the decision boundary plot shown in Figure \ref{fig:brand-attr-blind-scenario}. 
It can be seen that all samples lie in Apple's decision region, which indicates that the unknown samples are classified correctly. 

\begin{figure}[tb]
    \centering
    \includesvg[width=0.8\linewidth]{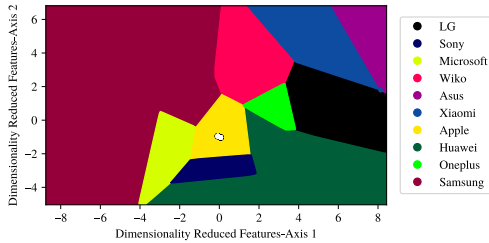}
    \caption{Example of blind device attribution scenario. The projected samples from the unknown device are shown with white markers with black contour.}
    \label{fig:brand-attr-blind-scenario}
\end{figure}

\subsection{Scenario 2: Manipulation Tool Identification}

The goal of this scenario is to determine the video editing tool used to manipulate a given video file.
We considered native video files from the acquisition devices as non-edited, and compare them with video files edited using Avidemux \rev{\cite{avidemux}}, Exiftool \rev{\cite{exiftool}}, ffmpeg \rev{\cite{ffmpeg}}, Kdenlive \rev{\cite{kdenlive}} and Premiere \rev{\cite{premiere}} as reported in \cite{yang_2020}.

In Table~\ref{table:software-identification}, we compare the F1-score of our method to previous work. 
Our method achieves a higher average F1-score compared to the previous work. 
The 2D feature distribution and decision boundary for this scenario are shown in Figure~\ref{fig:software-id-plot}.

\begin{table}[htpb]
	\tablesize
	\centering
	\caption{F1-score comparison of manipulation tool identification scenario.}
	Manipulation Tool Identification\\
	\begin{tabular}{rcc}
	\toprule
	\multicolumn{1}{c}{Tool} & Yang, \etal \cite{yang_2020} & Our Approach \\ \midrule
	Native &  0.97  &   1.00\\
	Avidemux &  0.99  & 0.98 \\
	Exiftool &  0.98  & 1.00 \\
	ffmpeg   &  0.94  & 1.00\\
	Kdenlive &  0.95  & 1.00\\
	Premiere &  1.00  & 0.99\\ \midrule
	Average & 0.97  &  0.99\\
	\bottomrule
	\end{tabular}
	\label{table:software-identification}
\end{table}

\begin{figure}[htpb]
    \centering
    \includesvg[width=0.8\linewidth]{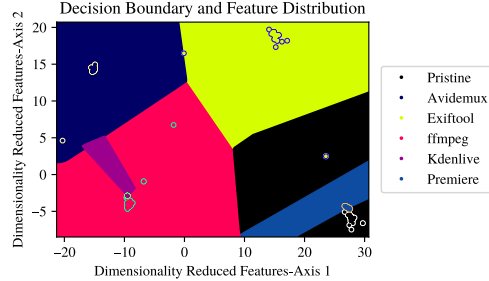}
    \caption{2D feature distribution and classification boundary for manipulation tool identification scenario.}
    \label{fig:software-id-plot}
\end{figure}

\subsection{Scenario 3: Social Network Attribution}

In our social network attribution scenario, we classify the source social network of the video files. 
If a video file is not downloaded from a social network or the video file comes from an unknown social network, it will be classified as ``other''. 
The F1-scores in this scenario are shown in Table \ref{table:social-network-f1}; the 2D feature distribution and decision boundary for this scenario are shown in Figure \ref{fig:social-network-plot}. 
For this task, our approach achieves high average F1-score of 0.99. 
The high performance also implies that each social network leaves a unique fingerprint on its videos.

\begin{figure}[htpb]
    \centering
    \includesvg[width=0.8\linewidth]{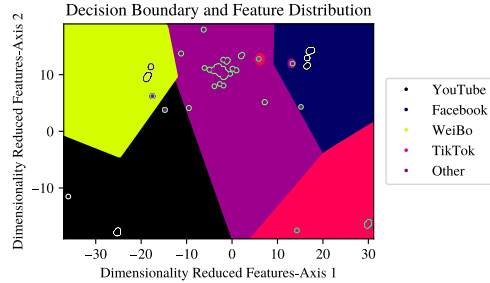}
    \caption{2D feature distribution and classification boundary for social network attribution scenario.}
    \label{fig:social-network-plot}
\end{figure}

\begin{table}[htpb]
    \tablesize
    \centering
    \caption{F1-score comparison of social network attribution scenario.}
    Social Network Attribution\\
    \begin{tabular}{rcc}
    \toprule
        \multicolumn{1}{c}{Social Network} & Yang, \etal \cite{yang_2020} & Our Approach \\ \midrule
        YouTube & 1.00 &  0.99\\
        Facebook & 1.00  &  1.00 \\
        WeiBo & 0.99 &  0.99\\
        TikTok &  1.00 &  1.00 \\
        Other   & - & 0.99 \\ 
        \bottomrule
    \end{tabular}
    \label{table:social-network-f1}
\end{table}

\subsection{Scenario 4: Manipulation Detection}

There are two sub-tasks within this scenario. 
As discussed above, for binary classification scenarios such as manipulation detection, LDA can only generate one-dimensional features. 
It limits the features' expression power after dimensionality reduction, which leads to inferior classification performance. 
Therefore, for binary classification problems, we prefer using a decision tree classifier without dimensionality reduction. 

\paragraph{Manipulation detection on videos files from social networks}
    
In this task, we detect manipulated videos given the fact that both pristine and edited videos are uploaded to social networks first. 
We compare the performance of using our features with different classification strategies to \cite{yang_2020}, and the results are shown in Table \ref{table:social-network-manip}. 
Using the EVA-7K dataset, the manipulation detection problem is unbalanced because there are much more edited videos than original ones (edited:pristine$\approx$9:1), meaning more samples with positive labels. 
In this case, the true positive rate (TPR) and the true negative rate (TNR) reflect a classifier's performance more objectively than F1-score.
It can be seen that our features combined with decision tree classifier achieve higher TNR for videos from all four social networks. 
When we use LDA and nearest neighbor classifier for this scenario, the classifier completely fails for videos from TikTok and WeiBo. 
It is likely because LDA can only generate one-dimensional features, which do not possess enough degrees of freedom to represent the complexity of this problem. 
Thus, the decision tree classifier is preferred for this scenario.

From Table \ref{table:social-network-manip}, we also have a glimpse of how each social network process uploaded videos. 
For WeiBo and TikTok videos, conducting further metadata-based forensic analysis becomes unreliable, which indicates they may have significantly altered videos uploaded to their platforms. 
YouTube videos can be classified perfectly, which implies that they apply minimum modification to videos' metadata.

\begin{table}[htpb]
\tablesize
\setlength{\tabcolsep}{3pt}
    \centering
    \caption{Performance evaluation metrics comparison between our approach and previous work. TPR and TNR stand for True Positive Rate and True Negative Rate, respectively. The accuracy score has been balanced.}
    Social Network Manipulation Detection\\
    \begin{tabular}{ccccc}
    \toprule
      &  & Yang, \etal \cite{yang_2020} & \makecell{Our Features+\\Decision Tree} & \makecell{Our Features+\\LDA \& NNC}\\ \midrule
    \multirow{3}{*}{\rotatebox{90}{\relsize{0.8} Facebook}} & Accuracy & 0.76 & 0.84 & 0.62\\
    & TNR & 0.40  & 0.87 &  0.30  \\
    & TPR & 0.86  & 0.82  & 0.95   \\[3pt] \midrule
    \multirow{3}{*}{\rotatebox{90}{\relsize{0.8} TikTok}} & Accuracy & 0.80 & 0.69 & 0.50\\
    & TNR &  0.51 & 0.94 &  0.00  \\
    & TPR &  0.75 & 0.43 &  1.00  \\ \midrule
    \multirow{3}{*}{\rotatebox{90}{\relsize{0.8} Weibo}} & Accuracy & 0.79 & 0.63 & 0.50\\
    & TNR & 0.45  & 0.57 &  0.00  \\
    & TPR & 0.82  & 0.68 &  1.00  \\ \midrule
    \multirow{3}{*}{\rotatebox{90}{\relsize{0.8} YouTube}} & Accuracy & 0.60 & 1.00 & 1.00\\
    & TNR & 0.36  & 1.00 &  1.00  \\
    & TPR & 0.74  & 1.00 &  1.00  \\ \bottomrule
    \end{tabular}
    \label{table:social-network-manip}
\end{table}

\paragraph{Manipulation detection on local videos}

In this task, we classify pristine videos and edited videos that are not exchanged via social network. 
To mimic the real world classification scenario, we employ a similar leave-one-out validation strategy as introduced in \cite{yang_2020}.
This approach takes the video files from one device model out as the validation set at a time. 
Since there is only one Microsoft device among 35 models, it is discarded in this scenario, {as described in \cite{yang_2020}}. {Because the Microsoft device either belongs to the training set or validation set, we are left with no samples to validate or no data to train.}
The mean balanced accuracy comparison of the 34-fold cross validation is shown in Table \ref{table:manip-detection-comparision}. 
Our approach achieves higher performance compared to previous works.

\begin{table}[htpb]
	\tablesize
    \centering
    \caption{Comparison of our method with previous works. The balanced accuracy is averaged over 34 folds.}
    Manipulation Detection on Local Videos\\
    \begin{tabular}{rc}
    \toprule
         &  Balanced Accuracy\\ \midrule
        G\"{u}era, \etal \cite{Guera2019_ICMLW} & 0.67\\
        Iuliani, \etal \cite{iuliani_2019} & 0.85\\
        Yang, \etal \cite{yang_2020} & 0.98\\
        Our Features + Decision Tree & 0.99\\ \bottomrule
    \end{tabular}
    \label{table:manip-detection-comparision}
\end{table}

\section{Conclusion}


In this paper, we proposed a video forensics approach for MP4 video files based on metadata embedded in video files. 
Our improved metadata extraction and feature representation scheme allows one to represent more metadata information in a compact feature vector. 
We use feature selection, dimensionality reduction, and nearest neighbor classification techniques to form interpretable and reliable decision rules for multiple video forensics scenarios. 
Our approach achieves better performance than other methods. 

{The performance of our method in many of the scenarios indicates that we need to increase our video forensics dataset to include more difficult cases.}
Our research also exposed the limitation of metadata-based forensics methods, namely its failure to analyze videos from specific social networks such as TikTok and WeiBo. 
This is a significant disadvantage compared to pixel-based methods. 
In the future, we plan to continue exploring the potential of metadata-based video forensics by adding the ability to parse more manufacturer-specific data models (e.g., Canon's CNTH tags \cite{canon_cnth}) and by looking into lower-level metadata in the distribution of audio/video samples as well as location of key frames in the video stream. 
We hope that metadata-based video forensics methods can be proved to be reliable in more forensic scenarios.

\section{Acknowledgment}


This material is based on research sponsored by DARPA and Air Force Research Laboratory (AFRL) under agreement number FA8750-20-2-1004. 
The U.S. Government is authorized to reproduce and distribute reprints for Governmental purposes notwithstanding any copyright notation thereon. 
The views and conclusions contained herein are those of the authors and should not be interpreted as necessarily representing the official policies or endorsements, either expressed or implied, of DARPA and Air Force Research Laboratory (AFRL) or the U.S. Government.
Address all correspondence to Edward J. Delp, \url{ace@ecn.purdue.edu}.

\bibliographystyle{ieee_fullname}
{
\apptocmd{\sloppy}{\hbadness 10000\relax}{}{}
\interlinepenalty=10000
\bibliography{refs}
}

\end{document}